\documentclass[a4paper,11pt]{article}
\pdfoutput=1 % if your are submitting a pdflatex (i.e. if you have
\usepackage{jheppub} % for details on the use of the package, please
\usepackage[T1]{fontenc} % if needed

\usepackage{graphicx}
\usepackage{amsmath}
\usepackage{amssymb}
\usepackage{hyperref}
\usepackage{textgreek}
\usepackage[ngerman, american]{babel}
\usepackage{booktabs}
\usepackage[version=4]{mhchem}
\usepackage[labelformat=simple]{subcaption}
\usepackage{siunitx}
\usepackage{tikz}
	\usetikzlibrary{calc}
	\usetikzlibrary{positioning}
\DeclareSIUnit\year{yr}

\newcommand{\cefe}{$\ce{^{55}Fe}$~}

\newcommand{\dd}{\mathop{}\!{\mathrm{d}}}

\title{\boldmath Search for solar axions produced through the axion-electron coupling \(g_{ae}\)
  using a new GridPix detector at CAST}

\emailAdd{phd@vindaar.de}
\emailAdd{schiffer@physik.uni-bonn.de}

\author[a]{K. Altenmüller,}
\author[b]{V. Anastassopoulos,}
\author[c]{S. Arguedas-Cuendis,}
\author[d]{S. Aune,}
\author[e]{J. Baier,}
\author[c]{K. Barth,}
\author[f,1]{\textdagger H. Bräuninger\note{Deceased},}
\author[g]{G. Cantatore,}
\author[c,h]{F. Caspers,}
\author[a]{J.F. Castel,}
\author[i]{S.A. Çetin,}
\author[j]{F. Christensen,}
\author[k,a]{C. Cogollos,}
\author[a]{T. Dafni,}
\author[c]{M. Davenport,}
\author[l]{T.A. Decker,}
\author[m]{K. Desch,}
\author[a]{D. Díez-Ibáñez,}
\author[c]{B. Döbrich,}
\author[d]{E. Ferrer-Ribas,}
\author[e]{H. Fischer,}
\author[c]{W. Funk,}
\author[a]{J. Galán,}
\author[a]{J.A. García,}
\author[n]{A. Gardikiotis,}
\author[d]{I. Giomataris,}
\author[c,o]{J. Golm,}
\author[p]{C.H. Hailey,}
\author[q]{M.D. Hasinoff,}
\author[r]{D.H.H. Hoffmann,}
\author[a]{I.G. Irastorza,}
\author[e]{J. Jacoby,}
\author[j]{A.C. Jakobsen,}
\author[s]{K. Jakovčić,}
\author[m]{J. Kaminski,}
\author[g,t]{M. Karuza,}
\author[c]{S. Kostoglou,}
\author[m,~now~u]{C. Krieger,}
\author[s,2]{\textdagger B. Lakić\note{Deceased},}
\author[c]{J.M. Laurent,}
\author[a]{G. Luzón,}
\author[c]{C. Malbrunot,}
\author[a]{C. Margalejo,}
\author[u]{M. Maroudas,}
\author[v]{L. Miceli,}
\author[a]{H. Mirallas,}
\author[w]{P. Navarro,}
\author[a]{L. Obis,}
\author[i,x]{A. Özbey,}
\author[i,y]{K. Özbozduman,}
\author[d]{T. Papaevangelou,}
\author[a]{O. Pérez,}
\author[l]{M.J. Pivovaroff,}
\author[z]{M. Rosu,}
\author[a]{E. Ruiz-Chóliz,}
\author[l,a]{J. Ruz,}
\author[m,3]{T. Schiffer\note{Corresponding author: schiffer@physik.uni-bonn.de},}
\author[m,4]{S. Schmidt\note{Corresponding author: phd@vindaar.de},}
\author[e]{M. Schumann,}
\author[v,aa]{Y.K. Semertzidis,}
\author[ab]{S.K. Solanki,}
\author[c]{L. Stewart,}
\author[c]{T. Vafeiadis,}
\author[l,a]{J.K. Vogel,}
\author[c,b]{K. Zioutas}

\affiliation[a]{Centro de Astropartículas y Física de Altas Energías (CAPA) \& Departamento de Física Teórica, University de Zaragoza, 50009 — Zaragoza, Spain}
\affiliation[b]{Physics Department, University of Patras, Patras, Greece}
\affiliation[c]{European Organization for Nuclear Research (CERN), 1211 Geneva 23, Switzerland}
\affiliation[d]{IRFU, CEA, Université Paris-Saclay, 91191 Gif-sur-Yvette, France}
\affiliation[e]{Physikalisches Institut, Albert-Ludwigs-Universität Freiburg, 79104 Freiburg, Germany}
\affiliation[f]{Max-Planck-Institut für Extraterrestrische Physik, Garching, Germany}
\affiliation[g]{University of Trieste and Instituto Nazionale di Fisica Nucleare (INFN), Sezione di Trieste, Trieste, Italy}
\affiliation[h]{European Scientific Institute, Archamps, France}
\affiliation[i]{Istinye University, Institute of Sciences, 34396, Sariyer, Istanbul, Turkey}
\affiliation[j]{DTU Space, National Space Institute, Technical University of Denmark, 2800 Lyngby, Denmark}
\affiliation[k]{Institut de Ciències del Cosmos, Universitat de Barcelona (UB-IEEC), Barcelona, Catalonia, Spain}
\affiliation[l]{Lawrence Livermore National Laboratory, Livermore, California 94550, USA}
\affiliation[m]{Physikalisches Institut, University of Bonn, 53115 Bonn, Germany}
\affiliation[n]{Institute of Quantum Computing and Quantum Technology NCSR “Demokritos", Athens, Greece}
\affiliation[o]{Institute for Optics and Quantum Electronics, Friedrich Schiller University Jena, Jena, Germany}
\affiliation[p]{Physics Department and Columbia Astrophysics Laboratory, Columbia University, New York, New York 10027, USA}
\affiliation[q]{Department of Physics and Astronomy, University of British Columbia, Vancouver, Canada}
\affiliation[r]{Xi'An Jiaotong University, School of Science, Xi'An, 710049, China}
\affiliation[s]{Rudjer Bošković Institute, Zagreb, Croatia}
\affiliation[t]{Faculty of Physics and Center for Micro and Nano Sciences and Technologies, University of Rijeka, 51000 Rijeka, Croatia}
\affiliation[u]{Universität Hamburg, Hamburg, Germany}
\affiliation[v]{Center for Axion and Precision Physics Research, Institute for Basic Science (IBS), Daejeon 34141, Republic of Korea}
\affiliation[w]{Department of Information and Communications Technologies, Technical University of Cartagena, 30203 — Murcia, Spain}
\affiliation[x]{Istanbul University-Cerrahpasa, Vocational School of Technical Sciences, 34320 Avcılar, Istanbul, Turkey}
\affiliation[y]{Boğaziçi University, Physics Department, Bebek, Istanbul, Turkey}
\affiliation[z]{Extreme Light Infrastructure - Nuclear Physics (ELI-NP), 077125 Magurele, Romania}
\affiliation[aa]{Department of Physics, Korea Advanced Institute of Science and Technology (KAIST), Daejeon 34141, Republic of Korea}
\affiliation[ab]{Max-Planck-Institut für Sonnensystemforschung, 37077 Göttingen, Germany}

\collaboration{CAST Collaboration}

\abstract{We present a search for solar axions produced through the axion-electron coupling
  (\(g_{ae}\)) using data from a novel 7-GridPix detector installed at
  the CERN Axion Solar Telescope (CAST). The detector, featuring
  ultra-thin silicon nitride windows and multiple veto systems,
  collected approximately 160 hours of solar tracking data between
  2017-2018. Using machine learning techniques and the veto systems,
  we achieved a background rate of
  $\SI{1.06e-5}{keV^{-1}.cm^{-2}.s^{-1}}$ at a signal efficiency of
  about $\SI{80}{\%}$ in the $\SIrange{0.2}{8}{keV}$ range. Analysis
  of the data yielded no significant excess above background, allowing
  us to set a new upper limit on the product of the axion-electron and
  axion-photon couplings of
  $g_{ae} \cdot g_{a\gamma} < \SI{7.35e-23}{GeV^{-1}}$ at 95\,\%
  confidence level for axion masses below $10\,\text{meV}$. This result improves upon the previous best
  helioscope limit and demonstrates the potential of GridPix
  technology for rare event searches. Additionally, we derived a limit
  on the axion-photon coupling of
  $g_{a\gamma} < \SI{9.0e-11}{GeV^{-1}}$ at 95\,\% CL, which,
  while not surpassing CAST's best limit, provides complementary
  constraints on axion models.}

\begin{document}
\maketitle
\flushbottom

\tableofcontents

\section{Introduction}
\label{sec:org1d8ba78}

The strong CP-problem \cite{PhysRevD.11.3583,tHooftU1,hooft1986instantons}
has a solution in the form of the QCD axion
\cite{PecceiQuinn1977_1,PecceiQuinn1977_2,AxionWeinberg,AxionWilczek},
which arises as a pseudo Nambu-Goldstone boson if a new scalar field
\(\alpha\) is added to the Standard Model Lagrangian. This promotes
the \(\theta\) parameter of the $\theta$-term

\[
\mathcal{L}_\theta = \theta \frac{g_\text{s}^2}{32\pi^2} G^{\mu\nu}_a \tilde{G}_{a\mu\nu},
\]
to a quantum field and leads to an effective coupling of the axion
to photons via a quark-gluon loop. As the core of the Sun provides an
environment of high rates of photon interactions, axions are expected
to be produced there following a spectrum similar to that of blackbody
radiation, with a rate proportional to \(g^2_{a\gamma}\). These axions
are searched for in axion helioscope experiments like the
\textbf{C}ERN \textbf{A}xion \textbf{S}olar \textbf{T}elescope (CAST)
\cite{ZIOUTAS1999480,PhysRevLett.94.121301} and in future the
\textbf{I}nternational \textbf{AX}ion \textbf{O}bservatory (IAXO)
\cite{Irastorza_2011,vogel2013iaxo,Armengaud_2014}.

In the DFSZ \cite{Zhitnitsky_DFSZ,DINE_DFSZ} invisible axion model, a
tree level coupling of the axion to electrons \(g_{ae}\) is present
\(\mathcal{L} \supset g_{ae} \frac{\partial_{\mu} a}{2 m_\text{e}}
\overline{\psi}_\text{e} \gamma^5 \gamma^{\text{\textmu}} \psi_\text{e}\). It can be shown
\cite{Redondo_2013} that this coupling can produce a dominant
contribution---depending on the choice of coupling parameter
values---to the solar axion flux by Compton, Bremsstrahlung and
photon-nucleus interactions in the Sun. In contrast to the
axion-photon flux with a peak energy at around \(\SI{3}{keV}\) the
axion-electron contributions peak at only around \(\SI{1}{keV}\). See
figure~\ref{fig:solar_axion_flux} for comparison of the two spectra.

The possible parameter range for $g_{ae}$ has been previously limited
by experiments. Some experimental approaches are sensitive exclusively to
the axion-electron coupling, while others rely on the axion-photon
coupling either for production or reconversion. A recent summary of
the theory of direct axion-electron sensitive experiments is found in
\cite{Berlin_2024}, which also includes an overview of the current
best limits in different mass ranges. The strongest direct detection
limit over all axion masses of interest comes from XenonNT
\cite{xenonNT_2022}. From astrophysical processes the most stringent
limits of $g_{ae}$ over a wide range of axion masses can be obtained
from the brightness of tip of the red-giant branch (TRGB) stars. Axion
production would induce more cooling, leading to a larger core mass at
helium burning ignition and thus brighter TRGB stars. Such limits are
of the order of
$g_{ae} < \num{1.3e-13} \text{ at } \SI{95}{\%} \text{
  CL}$. \cite{capozzi20_axion_neutr_bound_improv_with,straniero20_rgb_tip_galac_globul_clust}

For narrow mass ranges much tighter limits can be set with different
approaches. See also \cite{ciaran_o_hare_2020_3932430} for a very
detailed overview of the current best limits.

Helioscope experiments are only sensitive to the product of
$g_{ae} \cdot g_{a\gamma}$, due to the reliance on magnets to
reconvert the solar axions back into photons in the X-ray
regime. However, they provide limits for a wide range of axion
masses up to 10 meV. The current best such limit is
\(g_{ae} \cdot g_{a\gamma} < \SI{8.1e-23}{GeV^{-1}}\) from CAST in
2013, \cite{Barth_2013}.

In this paper we will present a new search for axions produced via the
axion-electron coupling $g_{ae}$ based on data taken at CAST in 2017
and 2018 using a 7-GridPix
\cite{medipix,campbell2005detection,CHEFDEVILLE2006490,vanderGraaf:2007zz,krieger13_gridpix_det, krieger2017gridpix}
detector for X-ray detection as the readout. First we will introduce
the detector in section~\ref{sec:detector}. Afterwards, we will
give an overview of the data taking campaign, section~\ref{sec:cast},
and then move to the data analysis and background reduction techniques
employed in section~\ref{sec:analysis}. This leads to the limit calculation
methodology used in section~\ref{sec:limit}. The result of the limit calculation
is then presented in section~\ref{sec:data_unblinding}, before we
conclude in section~\ref{sec:summary}.

For more details about this work, see \cite{schmidtPhD}.

\section{CAST setup and detector}
\label{sec:detector}

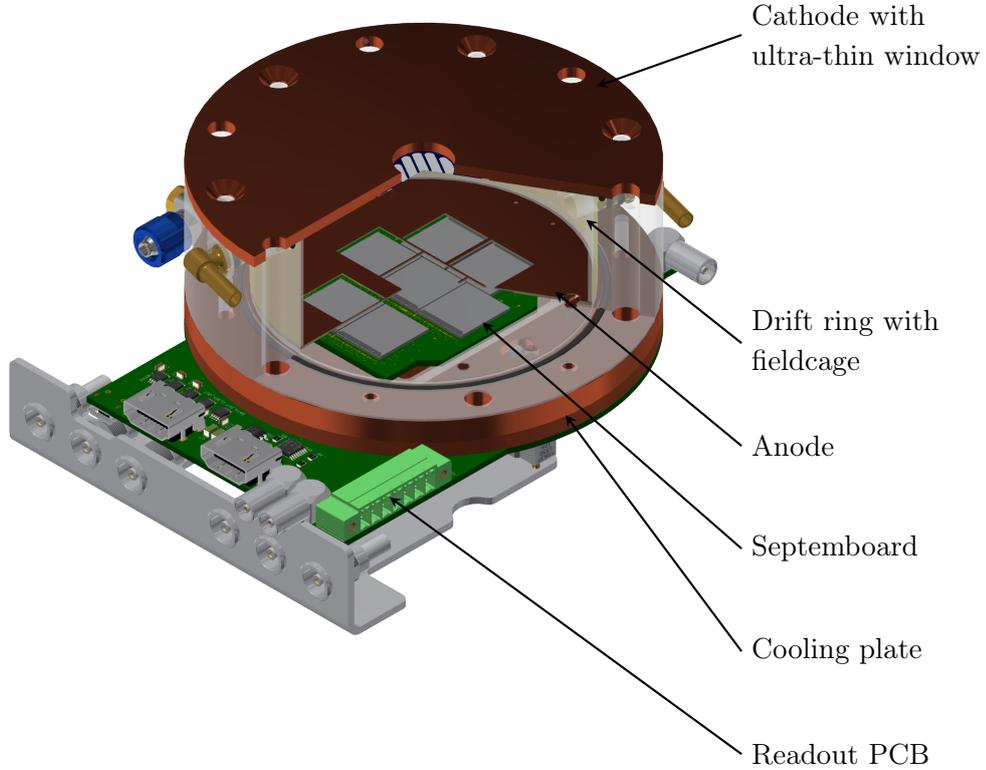
\begin{figure}[tp]
  \centering
  \begin{tikzpicture}
    \node[anchor=south west,inner sep=0] (Bild) at (0,0) {\includegraphics[width=0.9\textwidth]{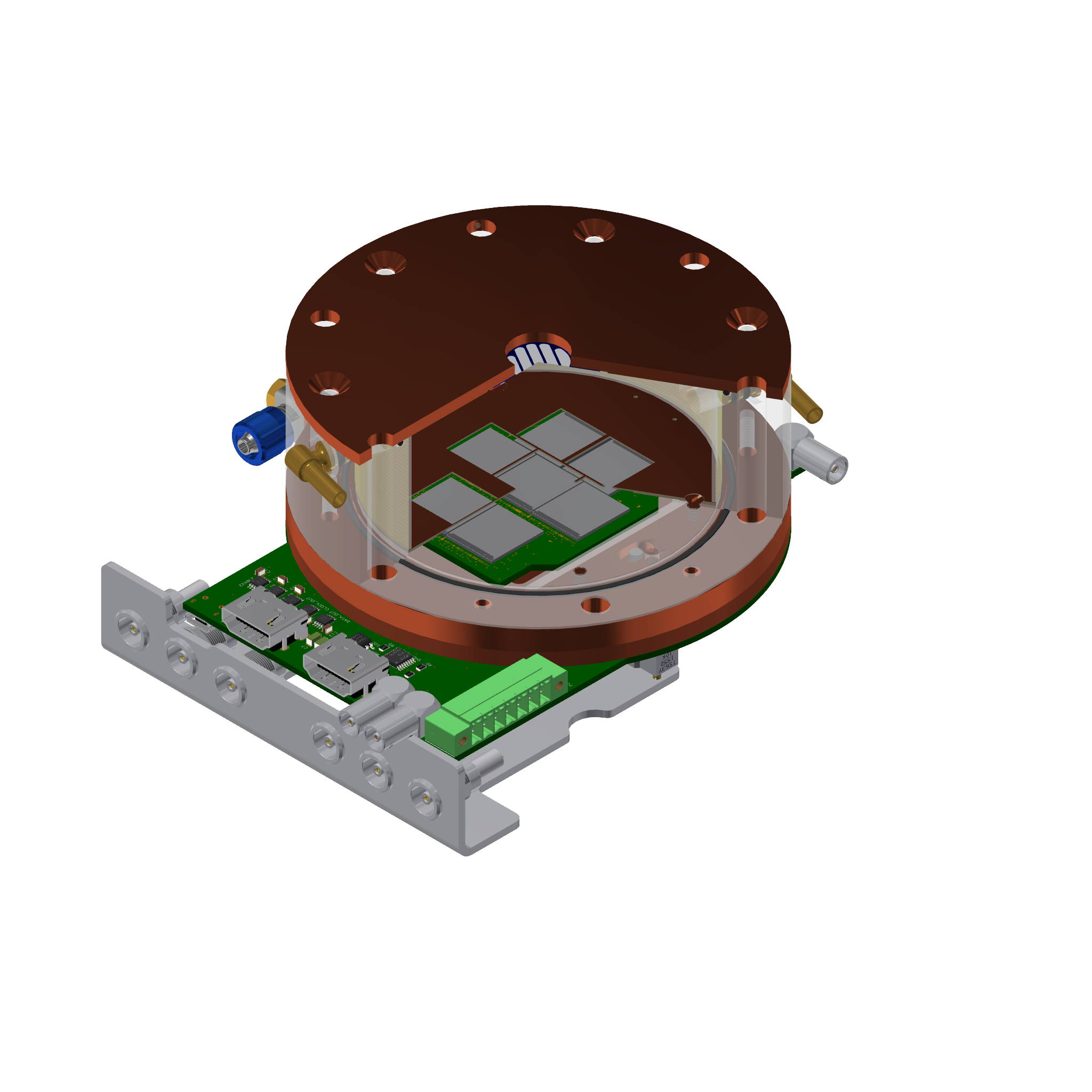}};
    \begin{scope}[x=(Bild.south east),y=(Bild.north west)]
      \node [right, align=left] (cathode) at (0.8,0.8){Cathode with\\ ultra-thin window};
      \draw [-stealth, thick, black](cathode.west) -- (0.66,0.75);
      \node [right, align=left] (driftring) at (0.8,0.5){Drift ring with\\ fieldcage};
      \draw [-stealth, thick, black](driftring.west) -- (0.65,0.62);
      \node [right, align=left] (anode) at (0.8,0.4){Anode};
      \draw [-stealth, thick, black](anode.west) -- (0.62,0.55);
      \node [right] (sep) at (0.8,0.3){Septemboard};
      \draw [-stealth, thick, black](sep.west) -- (0.55,0.52);
      \node [right] (cooling) at (0.8,0.2){Cooling plate};
      \draw [-stealth, thick, black](cooling.west) -- (0.63,0.43);
      \node [right] (read) at (0.8,0.1){Readout PCB};
      \draw [-stealth, thick, black](read.west) -- (0.46,0.35);
    \end{scope}
  \end{tikzpicture}
  \caption{\label{fig:detector:cut}Render of the 7-GridPix detector,
    cut open. Visible is the layout of the 'Septemboard', with seven GridPixes)
    as the readout. Not visible is a small SiPM-based scintillator installed below the PCB.}
\end{figure}

\begin{figure}[tp]
\begin{subfigure}{0.5\linewidth}
  \includegraphics[width=1.0\linewidth]{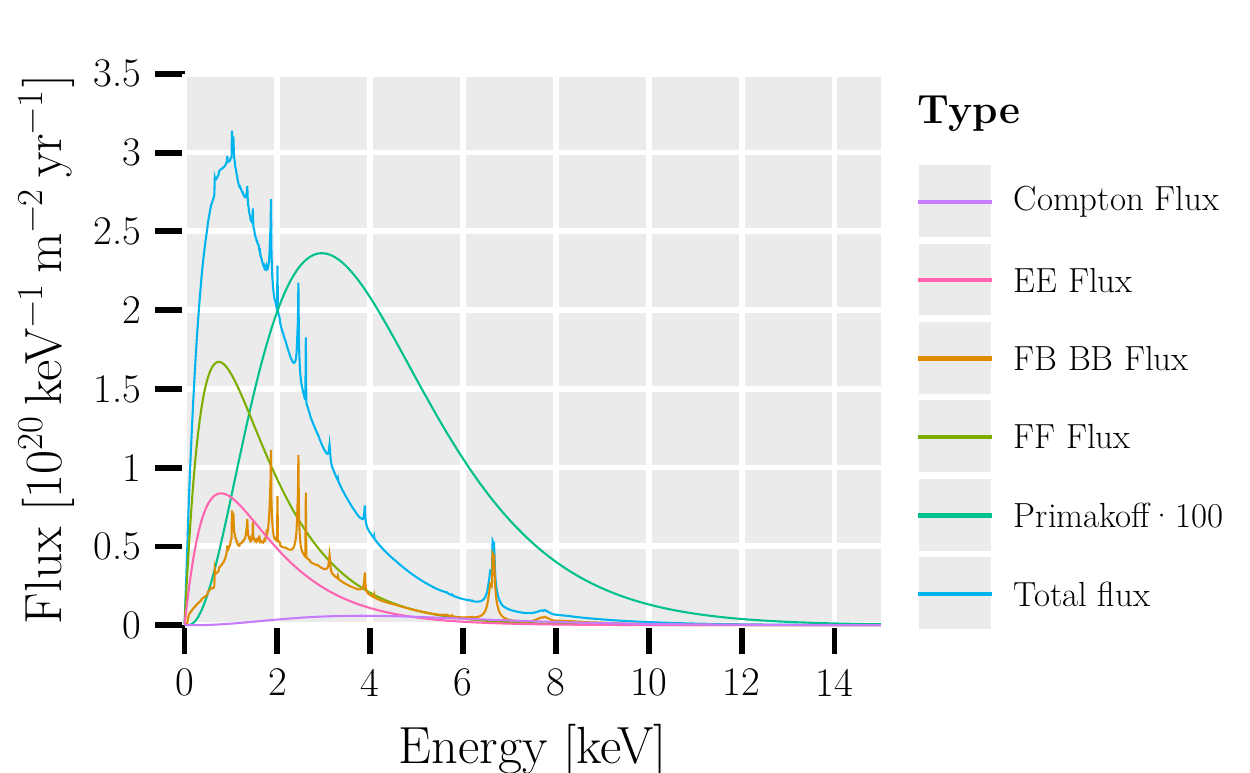}
  \caption{Differential solar axion flux}
  \label{fig:solar_axion_flux}
\end{subfigure}%
\begin{subfigure}{0.5\linewidth}
  \includegraphics[width=1.0\linewidth]{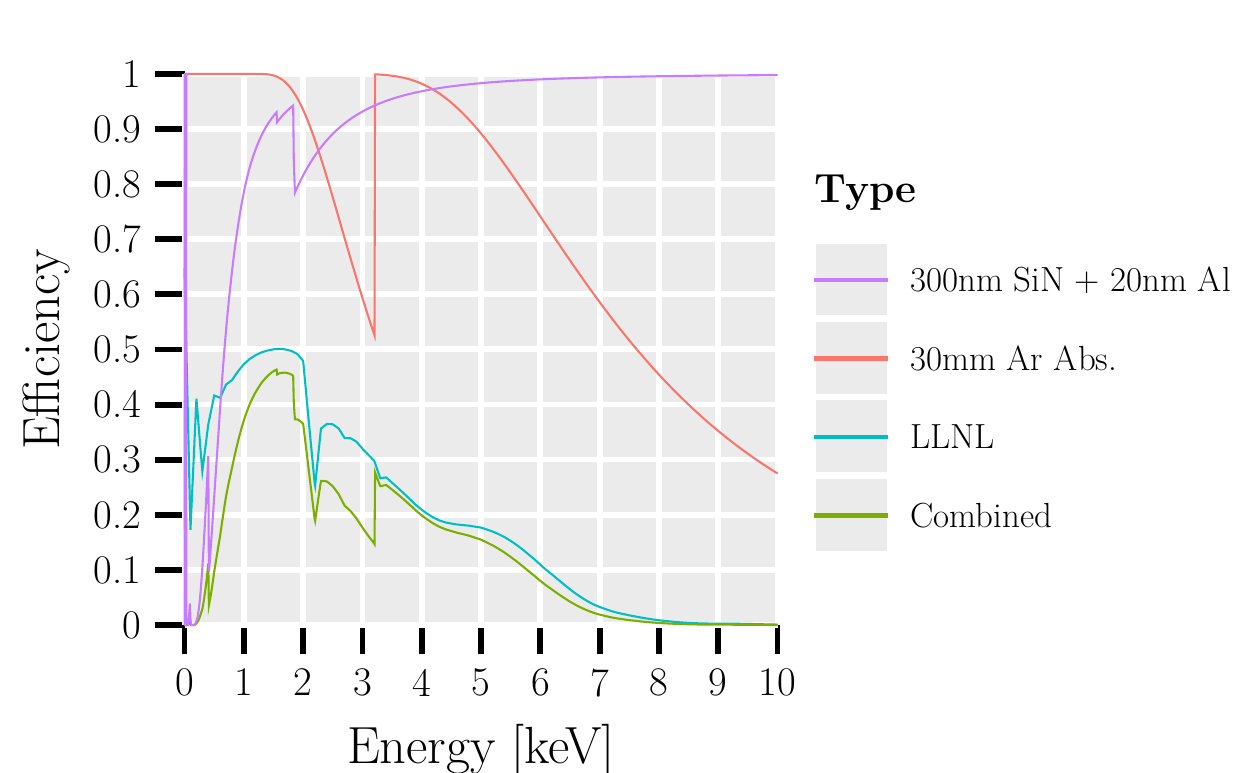}
  \caption{Detection efficiency}
  \label{fig:detector:detection_efficiency}
\end{subfigure}%
\caption{\subref{fig:solar_axion_flux}: Differential solar axion flux
  assuming a distance to the Sun of $\SI{0.989}{AU}$.
  'Primakoff$\cdot 100$' corresponds to the pure axion-photon
  $g_{a\gamma}$ production in the Sun for
  $g_{a\gamma} = \SI{1e-12}{GeV^{-1}}$, multiplied by a factor $100$
  to make this contribution visible for the choice of parameters. The
  other contributions are all via the axion-electron coupling at a
  value of $g_{ae} = \num{1e-13}$. 'EE': electron-electron
  bremsstrahlung, 'FB BB': free-bound, bound-bound interactions
  (axio-recombination and axio-deexcitation), 'FF': free-free
  transition (electron-ion bremsstrahlung).
  \subref{fig:detector:detection_efficiency}: Detection efficiency of
  the detector based on the window (\ce{Si N_4} and \ce{Al})
  transmission, gas absorption ('30\,mm Ar Abs.') and X-ray telescope
  effective area ('LLNL'). }
\label{fig:detector:flux_efficiency}
\end{figure}

The CAST experiment \cite{ZIOUTAS1999480,PhysRevLett.94.121301}
provides four experiment places to install detectors. In the last
years of operation only the side of the magnet pointing towards
sunrise were in operation, as there were X-ray optics behind each of
the two magnet bores of the LHC prototype dipole magnet.  One of these
optics was a custom built NuSTAR-like X-ray optic, consisting of
slumped glass layers, which are conical approximations \cite{Petre:85}
to the hyperbolic and parabolic mirrors of a Wolter type I optic.  It
was specifically designed for CAST, hence the effective area was
optimized in the energy range of interest for axion searches
\cite{llnl_telescope_first_cast_results}. The optic focuses the
produced X-rays onto a focal plane \(\SI{1500}{mm}\) behind the optic.
The GridPix detector used for the described
measurement was connected with a vacuum system to the optic such that
the focal plane of the optic is located in the center of the gas
volume. A render of the detector can be seen in
figure~\ref{fig:detector:cut}. It uses a similar design in terms of
dimensions as the MicroMegas (MM) detector used in \cite{cast_nature}. A
\(\SI{78}{mm}\) diameter acrylic glass housing with an inner height of
\(\SI{30}{mm}\) to contain a gas mixture of Argon/Isobutane
\(\SI{97.7}{\%}/\SI{2.3}{\%}\) at \(\SI{1050}{mbar(a)}\) chamber
pressure. As a barrier between the gas and the vacuum, a
\(\SI{300}{nm}\) thick silicon-nitride \(\ce{Si N_x}\) window is
employed to provide X-ray transmission while withstanding the
\(\SI{1050}{mbar}\) pressure
difference. Figure~\ref{fig:detector:detection_efficiency} shows the
transmission efficiency of the window as well as the absorption efficiency of
X-rays in the energy range of interest. The usage of
ultra-thin \ce{Si N_x} windows improves the X-ray transmission in the interesting range from $0\text{--}3\,\text{keV}$
by $8.13\,\%$ compared to commonly used $\SI{2}{\micro \meter}$ Mylar windows
in the energy range of interest for searches for
\(g_{ae}\). Figure~\ref{fig:solar_axion_flux} shows the differential
solar axion flux expected when taking the axion-electron coupling into
account. The used coupling constants are $g_{ae} = \num{1e-13}$ and
$g_{a\gamma} = \SI{1e-12}{GeV^{-1}}$.

\begin{figure}[tp]
\centering
\includegraphics[width=.9\linewidth]{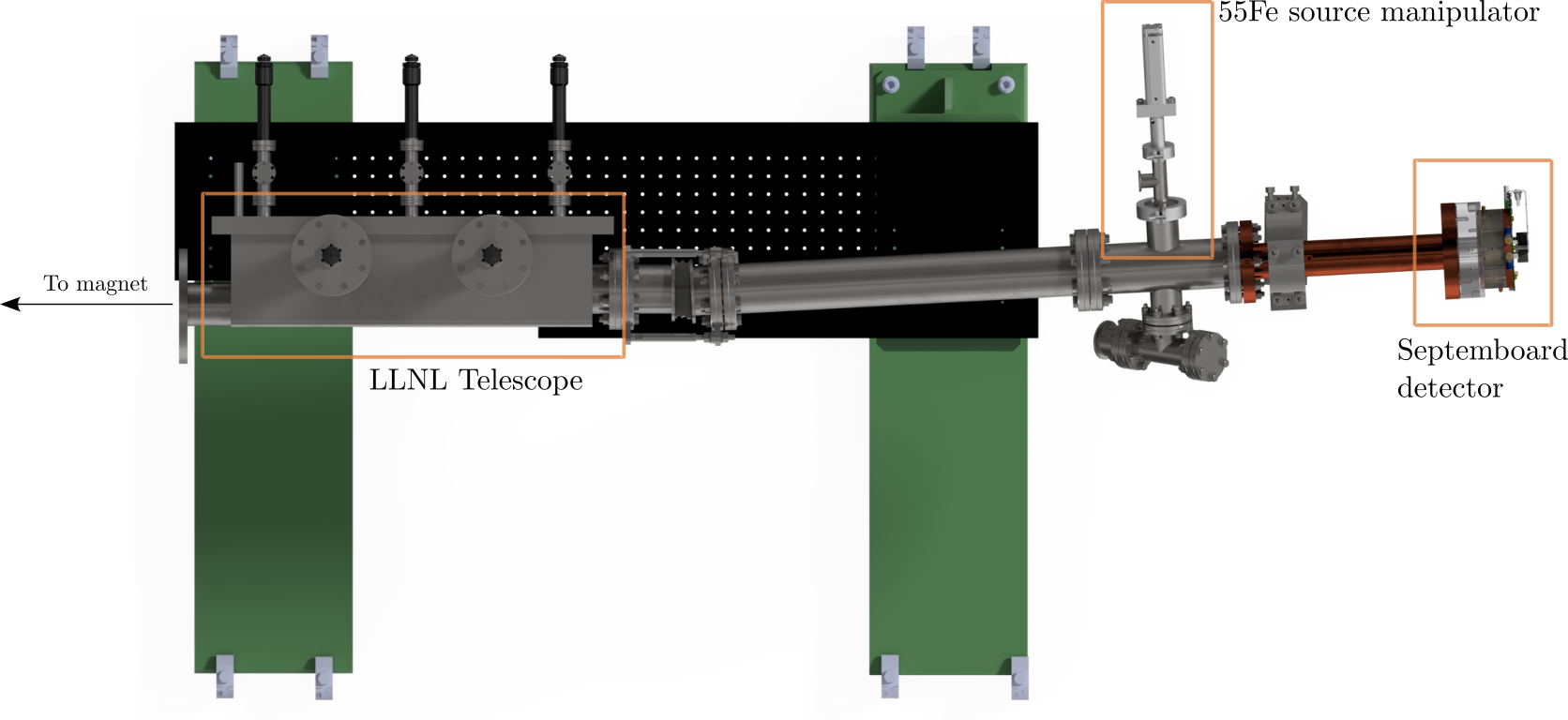}
\caption{\label{fig:cast:render_beamline_setup}Render of the detector
  setup up to the magnet end cap as seen from above. The beamline
  kinks away from the other beamline ("below" in this image) to
  provide more space for two detectors at the same time.}
\end{figure}

\begin{figure}[tp]
\centering
\includegraphics[width=.9\linewidth]{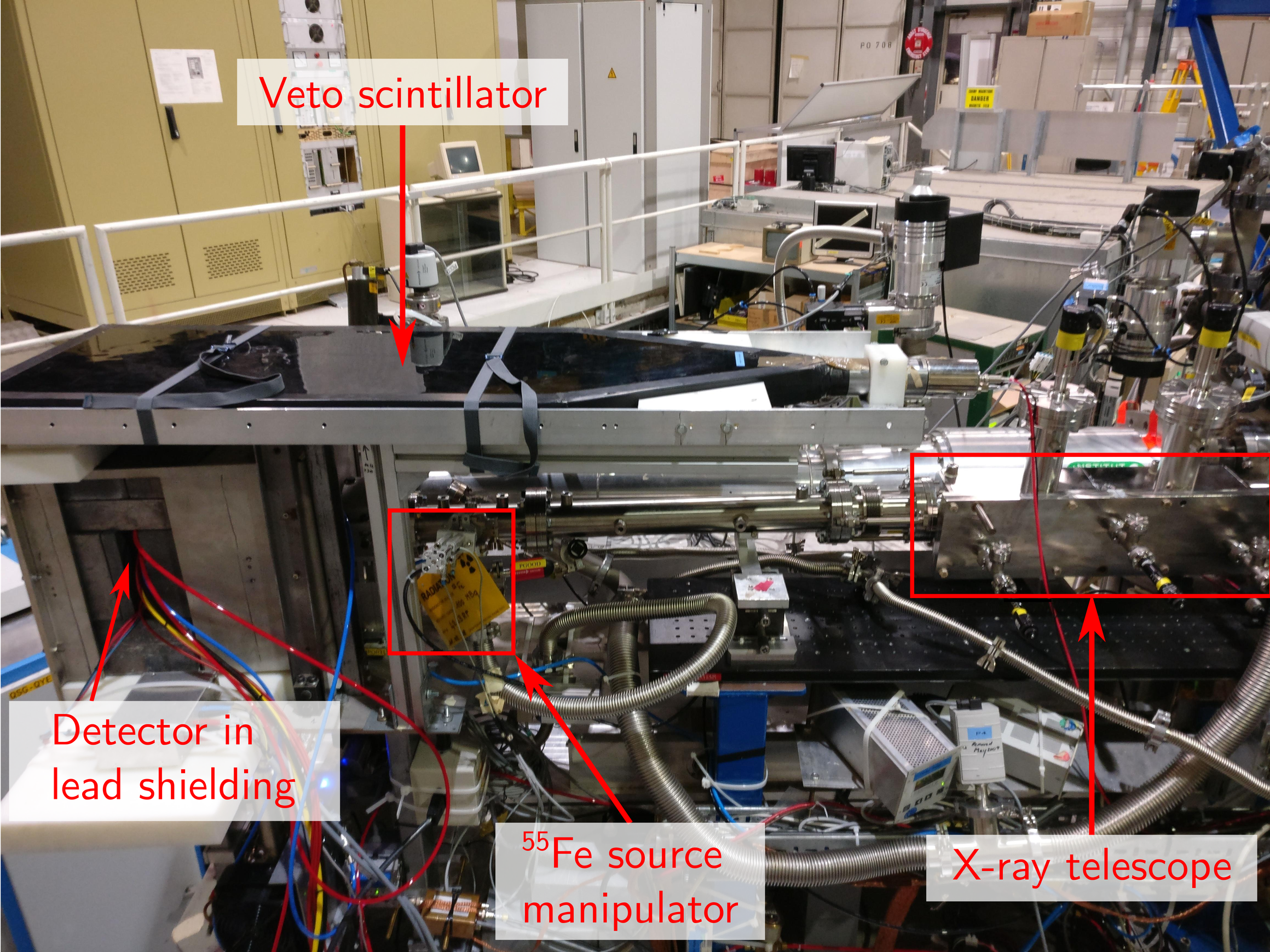}
\caption{\label{fig:cast:annotated_setup}Annotated setup as installed
  in October 2017 for the first data taking campaign. The detector is
  housed within its lead shielding, with the veto scintillator covering a
  large angular portion above the detector. The \cefe source
  manipulator is seen head-on. On the right towards the magnet we
  see the housing of the X-ray telescope.}
\end{figure}

A render of the setup is shown in
figure~\ref{fig:cast:render_beamline_setup} and an annotated image of
the real setup in figure~\ref{fig:cast:annotated_setup}.  Since the
expected image of the converted solar axions on the detector is of
great importance for the data analysis, a dedicated raytracing
simulation for X-rays was developed \cite{traxer} to simulate the
expected axion image, based on the expected axion emission in the Sun,
conversion probability in the magnet, detailed description of the
X-ray optics and detector. The raytracing code agrees with
measurements of the telescope taken at the PANTER test facility within
\(\SI{2.6}{\%}\) when comparing the half power diameter of three
different X-ray lines, $\ce{Al}$ K\textalpha~($\SI{1.49}{keV}$),
$\ce{Ti}$ K\textalpha~($\SI{4.51}{keV}$) and $\ce{Fe}$ K\textalpha
~($\SI{6.41}{keV}$).

The detector itself is a gas ionisation chamber using the GridPix
technology
\cite{medipix,campbell2005detection,CHEFDEVILLE2006490,vanderGraaf:2007zz}.
A GridPix consists of an integrated grid structure, as a gas
amplification stage, on top of a Timepix ASIC. The grid is produced in
a photolithographic process to perfectly align one hole with each
pixel. This enables both sufficient amplification and spatial
resolution to detect individual primary electrons produced in the
ionisation chamber. At the center of the detector is the 'Septemboard'---a PCB equipped with seven GridPixes.  The detector is aligned such
that the expected axion image is focused on the central GridPix. The
outer six GridPixes extend the sensitive area to allow for additional
background rejection. Because the seven GridPixes combined produce a
significant amount of heat due to thermal losses a custom-built water
cooling system is installed below the intermediate board. It is based
on an oxygen-free copper body with \(\SI{3}{mm}\) diameter channels
for water transport. This setup provides sufficient cooling for the
GridPixes operation.

The GridPixes used in this detector are based on the Timepix ASIC
\cite{LLOPART2007485_timepix}, which is read out in a frame based
fashion. The Timepix consists of \(256\times256\) pixels with
\(\SI{55}{\micro m}\) pitch and an active area of
\(\num{1.4}\times\SI{1.4}{\cm^2}\). Each pixel can either be used to
record Time-over-Threshold (ToT) values for charge accumulation or
Time-of-Arrival (ToA) information, but not both. For this reason an
additional flash ADC (FADC) is connected to the grid of the central
GridPix. It both acts as a readout trigger in case a fixed threshold
is exceeded while also providing further charge as well as
longitudinal time information. Reading out the seven GridPixes needs
\(\SI{175}{ms}\) due to the daisy-chained data output of the
GridPixes. Thus, short shutter times would lead to a large dead time
ratio.  Finally, two scintillators behind and above the detector are
installed to record potential muons traversing the detector.  To
reduce the amount of direct background from cosmics the detector sits
inside of a \(\SI{10}{cm}\) lead housing. A \cefe calibration source
is available at CAST and can be inserted into the beamline in front of
the detector using a pneumatic system. It provides a reference for
detector stability via daily calibration runs, for energies of
\(\SI{5.9}{keV}\) (photopeak) and \(\SI{3}{keV}\) (argon escape peak).

\section{Data taking}
\label{sec:cast}
The detector described in section~\ref{sec:detector} was first installed at
CAST in September 2017. The first data taking period started in Oct
2017 and went until April 2018 (Run-1). As CAST was typically shut
down over the summer months, the data taking continued from Oct 2018
to Dec 2018 (Run-2). During this data taking about \(\SI{3500}{h}\) of
background data and \(\SI{180}{h}\) of solar tracking data were recorded
(see table~\ref{tab:cast:summary}). The detector ran with a dead time
ratio of about \(\SI{10}{\%}\) thanks to a long shutter time of about
\(\SI{2.2}{s}\). Data was taken in ToT mode to later compute the recorded
charge of each pixel. Due to a firmware bug, scintillator triggers were
only saved in Run-2.

\begin{table}[tp]
\centering
\begin{tabular}{lrrrrr}
\toprule
 & Solar tracking [h] & Active s. [h] & Background [h] & Active b. [h] & Active [\%]\\[0pt]
\midrule
Run-1 & 106.01 & 93.37 & 2391.16 & 2144.12 & 89.65\\[0pt]
Run-2 & 74.30 & 67.01 & 1124.93 & 1012.68 & 90.02\\[0pt]
Total & 180.30 & 160.38 & 3516.09 & 3157.35 & 89.52\\[0pt]
\bottomrule
\end{tabular}
\caption{\label{tab:cast:summary}Overview of the total data taken with
  the Septemboard detector at CAST in the time between October 2017
  and December 2018. 'Active s.' and 'Active b.' refers to the total
  solar tracking and background time excluding the dead time due to
  readout of the detector.}

\end{table}

\section{Data analysis and background reduction}
\label{sec:analysis}
The data is read out from the detector using the Timepix Operating
Software (TOS) \cite{lupberger2016pixel,TOS_github}. Frames of hit
pixels are saved as ASCII files containing signal or background
clusters. Hit pixels originate typically from individual ionization
electrons produced by the interacting particles, spread out by
diffusion due to the drift in the gas volume. The files are parsed and
stored in HDF5 \cite{hdf5} files using the TimepixAnalysis framework
\cite{TPA}, which handles the entire analysis chain from raw data
processing through final limit calculation. The data reconstruction is
based on analyzing individual clusters of pixels. Expected signals
from X-rays are roughly spherical in shape whereas the dominant
background sources are less symmetrical (e.g. tracks from muons). The
background rejection algorithms are therefore based on geometric
properties of the clusters.

As part of the data reconstruction, first a distance based clustering
algorithm is used to reconstruct individual clusters from raw data. From
here, each cluster is handled separately. First the long and short
axes are determined via optimization of the maximum pixel-to-pixel
distance and its orthogonal extent. The long and short axes provide
the basis for a variety of different geometric properties. The
geometric properties derived from these axes, along with other cluster
characteristics, are shown in table~\ref{tab:geometric_properties}.

Next, the ToT values of each pixel are converted into a charge in
electrons. Based on \(\SI{90}{min}\) time intervals all charge values
recorded are accumulated to a histogram, from which the gas gain can
be extracted via a Pólya distribution fit (following
\cite{alkhazov1970statistics}). This calibration works with any
dataset---background or calibration---as it depends only on individual
pixel charge distributions. Calibration runs are used to correlate
potentially varying gas gains over time with a linear calibration
function mapping recorded charges to target energies of the \cefe
calibration source. This allows to compute the energy of any cluster
from its charge by using the associated gas gain from the
\(\SI{90}{min}\) interval it is part of.

With the data fully reconstructed and calibrated, each cluster needs to be
classified as either background- or signal-like. The method employed
to compute the final limit follows.

The initial classification is based on a machine learning approach
using a Multi-Layer Perceptron (MLP)
\cite{rosenblatt1958perceptron,amari67_mlp,schmidhuber22_history}
\footnote{Due to the rich and long history of artificial neural
  networks picking "a" or only a few citations is tricky. Rosenblatt
  \cite{rosenblatt1958perceptron} introduced the perceptron while
  Amari's work \cite{amari67_mlp} first combined a perceptron with a
  non-linear activation function and used gradient descent for
  training. See Schmidhuber's recent overview for a detailed history
  leading up to modern deep learning \cite{schmidhuber22_history}.}, a
fully connected feed-forward network. The specific network used has 14
input neurons, which correspond to twelve geometric properties of the
input cluster, its total charge and the gas diffusion coefficient
$D_\text{T}$ during the time this cluster was taken. All properties are shown
in table~\ref{tab:geometric_properties}. $D_\text{T}$ is determined by an
iterative optimization approach comparing the transverse RMS
distribution via the Cramér-von-Mises criterion
\cite{cramer28_gof,mises36_gof,anderson62_cvm_gof} of simulated data
with that of real data of a data taking run. The simulation is a Monte
Carlo code, which takes a diffusion coefficient $D_\text{T}$, gas gain and
target energy as an input. It samples a conversion point for an X-ray
and models active pixels via expected electron drift to the readout using
$D_\text{T}$. Each electron receives a charge based on sampling from a Pólya distribution
of the given gas gain. The transverse RMS of a cluster is directly
proportional to the transverse diffusion coefficient $D_\text{T}$. To
determine $D_\text{T}$ for a background run, we make use of the fact that
along the short axis of a cluster both background tracks and X-ray
clusters only undergo diffusion from a single line and point,
respectively.

\begin{table}[tp]
  \centering
  \begin{tabular}{ll}
    \toprule
    Property & Meaning\\
    \midrule
    Eccentricity & eccentricity of the cluster\\
    RMS \& Skewness \& Kurtosis & RMS/skewness/kurtosis along long/short axis\\
    Length \& Width & size along long/short axis\\
    LengthDivRmsTrans & length divided by transverse RMS\\
    RotationAngle & rotation angle of long axis over chip coordinate system\\
    FractionInTransverseRms & fraction of pixels within transverse RMS around center\\
    TotalCharge & integrated charge of total cluster in electrons\\
    \(D_\text{T}\) & transverse gas diffusion coefficient\\
    \bottomrule
  \end{tabular}
  \caption{\label{tab:geometric_properties}Table of all the properties
    of a single cluster used as input for the MLP. RMS, skewness and kurtosis
    are computed for each axis individually.}
\end{table}

\begin{figure}[tp]
\centering
\includegraphics[width=0.9\textwidth]{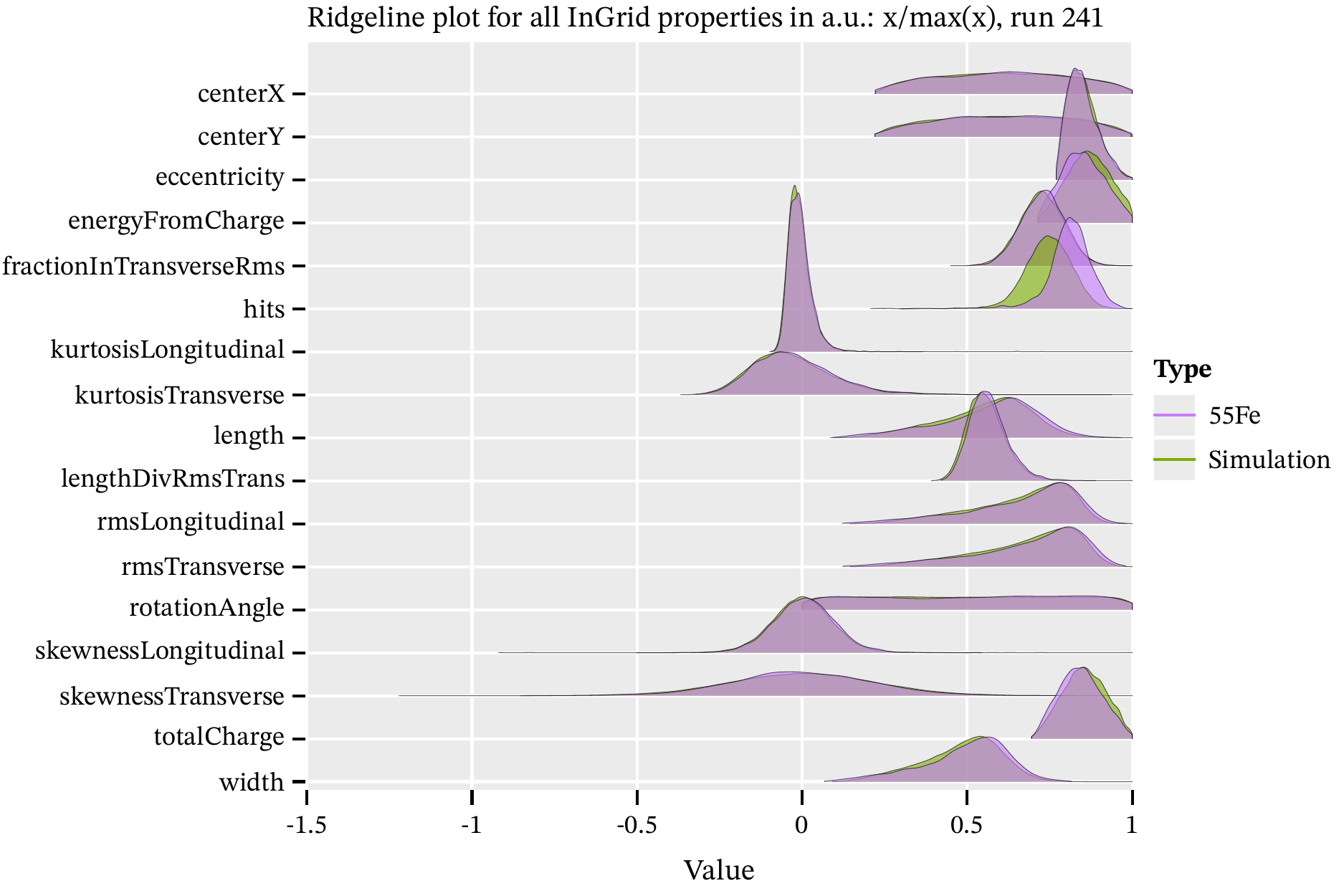}
\caption{\label{fig:analysis:comparison_sim_real} Comparison of all
  geometric properties of simulated and real data of the \cefe
  photopeak clusters of run 241 in a ridgeline plot. Each ridge shows
  a KDE of the data. The gas gain and gas diffusion were first
  extracted from the real run data to simulate events of the
  photopeak.  The two datasets agree very well with the exception of
  the number of hits (expected). }
\end{figure}

The MLP has two hidden layers with \(\num{30}\) neurons each and two
output neurons, one for X-ray-like and one for background-like
clusters. It uses \texttt{tanh} as an activation function and the mean
squared error (MSE) as a loss function. Training was run over
\(\num{83000}\) epochs of \(\num{250000}\) X-ray and background events
each. The background events are taken from the extensive set of
background data of the \emph{outer chips}, while the X-ray events are
fully simulated. It is the same event simulation as for the
determination of the gas diffusion. X-rays of a wide range of gas
diffusion coefficients, gas gains and uniform energies are
generated. The distributions of geometric properties between simulated
and real data with the same parameters agree very well, see figure
\ref{fig:analysis:comparison_sim_real}. The total number of hits per
cluster differs between simulation and real data, as the simulation
uses a simplified model for neighboring pixel activation due to UV
photons from the gas amplification. However, this does not affect the
MLP training, since it only depends on the derived geometric
properties which are well reproduced by the simulation. Based on the
Kolmogorov-Smirnov test statistic, the largest deviation across the
empirical distribution functions for simulated and real calibration
data is below \(\SI{5}{\%}\) when averaged over all properties (except
the 'hits' parameter) and all runs.

To differentiate signal from background events we define a cut on the
output value of one of the neurons (as the two output neurons are
complementary) by desiring a specific software efficiency for X-ray
data:

\[
\epsilon_\text{S} \equiv \epsilon_{\text{eff}} = \frac{\int_0^{x'} \text{MLP}(c) \, \mathrm{d} x}{\int_0^{\infty}
\text{MLP}(c)\, \mathrm{d} x} = \text{EDF}(x')
\]

where $\text{MLP}(c)$ corresponds to the MLP output for a given cluster $c$ and the
value \(x'\) then corresponds to the value to cut at. As
indicated this is computed based on the empirical cumulative
distribution function (EDF) in practice.

While the MLP cut value is determined using simulated events, the
actual signal efficiency used in the limit calculation is calibrated
using real X-ray data: both \cefe calibration data and measurements
at eight different energies from an X-ray generator. Different
energies are useful, because any model working on similar gaseous
detector data is inherently energy dependent. Lower energies mean
fewer activated pixels, which reduces the possible differentiation
between X-rays and background data.

The scintillators provide an additional veto: events are rejected if
the center GridPix shows activity within \(\SI{2.5}{\micro s}\) of a
scintillator trigger, using the FADC as the GridPix readout trigger
for X-ray energies above $1-2\,\text{keV}$.

For clusters including FADC data, this data is also used to veto some more events. It
effectively records the longitudinal shape information of the
clusters. This is determined by the event shape and the longitudinal
drift velocity and diffusion. These can be estimated from theory and
the \cefe calibration datasets provide further validation. Based on
this, all events outside the \(1^{\text{st}}\) and \(99^{\text{th}}\)
percentile of the FADC signal rise time distribution are
removed.

Finally, the outer ring of GridPixes is used as a veto. In case of an
X-ray like cluster on the center chip, the data of the outer chips is
considered. If a cluster on the outer chip is found whose long axis
points towards the center cluster, the event is vetoed.

The last veto comes with an efficiency penalty due to potential for
random coincidences. Shutter times are very long, yet only the center
chip has a trigger. There is the chance of uncorrelated activity on
the outer chip. To estimate this new events are bootstrapped from a
combination of center GridPix clusters that pass the MLP cut and data
from the outer chips of any other event. The rate of triggering this
veto then is purely random coincidence. Using this approach the
efficiency of this veto is estimated to \(\SI{86}{\%}\).

\begin{figure}[tp]
\centering
\includegraphics[width=0.9\textwidth]{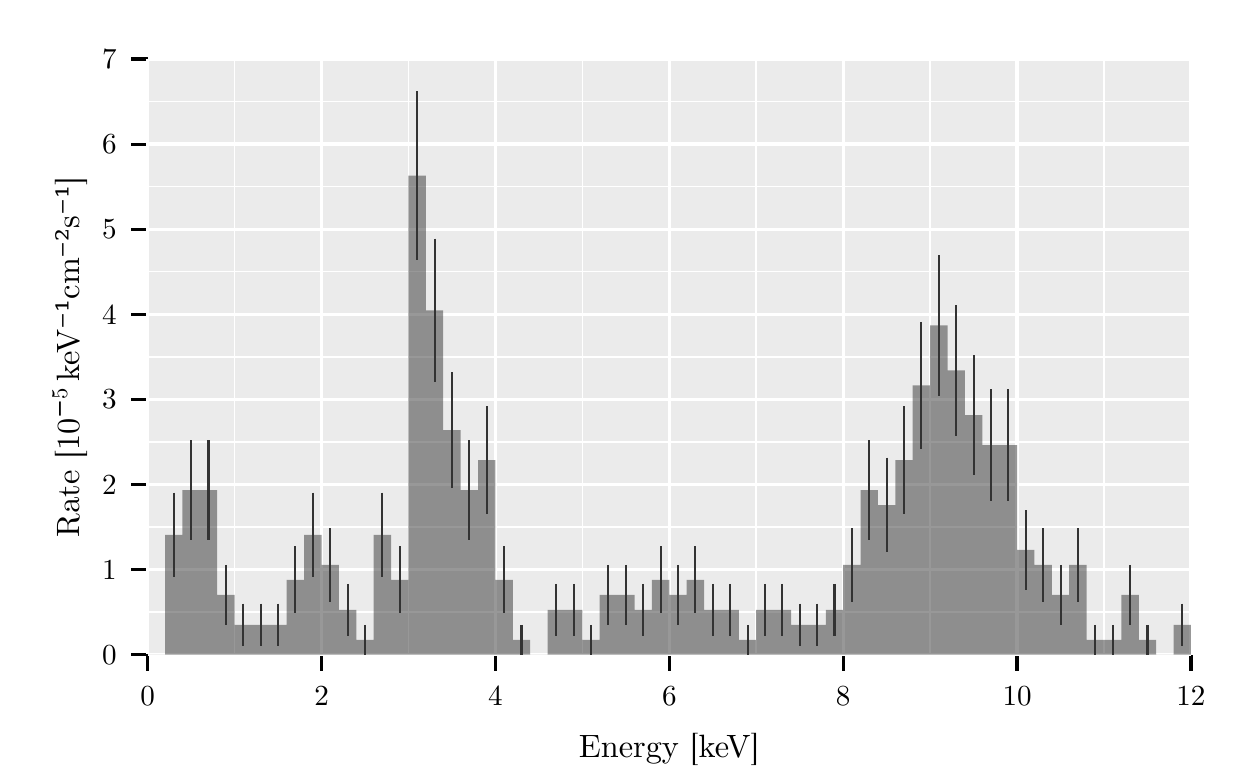}
\caption{\label{fig:analysis:background}Background rate in the center \(\num{5}\times\SI{5}{mm^2}\) using the MLP at \(\SI{95}{\%}\) effective efficiency and all mentioned vetoes. Over the energy range of \(\SIrange{0.2}{8}{keV}\) an average background rate of \(\SI{1.05552(6900)e-05}{keV^{-1}.cm^{-2}.s^{-1}}\) is achieved. The rate can be improved at the cost of signal efficiency.}
\end{figure}

Combining the MLP cut with all vetos, we achieve a background rate over the center
\(\num{5}\times\SI{5}{mm^2}\) area of the center chip (region of lowest
background) of \(\SI{1.05552(6900)e-05}{keV^{-1}.cm^{-2}.s^{-1}}\)
within \(\SIrange{0.2}{8}{keV}\), shown in
figure~\ref{fig:analysis:background}. This is for the case of an effective MLP
efficiency of \(\SI{95}{\%}\).

\section{Limit calculation method}
\label{sec:limit}
To compute a limit on the axion-electron coupling constant we use a
Bayesian approach based on finding the \(95^{\text{th}}\) percentile of
the marginal posterior likelihood. Our initial likelihood function is
derived from a ratio of two Poisson distributions, the signal plus
background hypothesis over the pure background hypothesis:

\[
\mathcal{L} = \prod_i \frac{P_{\text{pois}}(c_i; s_i + b_i)}{P_{\text{pois}}(c_i; b_i)}
\]

which runs over all bins \(i\) and \(c_i\) are the number of
candidates in each bin, \(s_i, b_i\) the signal and background,
respectively. The signal \(s_i\) is the expected amount of signal in bin
\(i\) based on the solar axion flux and all detection efficiencies
included. The background is given by a background model constructed
from the entire background dataset taken at CAST during non-tracking
times. This likelihood is taken to the unbinned likelihood by choosing bins in
time such that each bin only contains either 0 or 1 candidates.

The likelihood function simplifies to

\[
\mathcal{L} = e^{-s_{\text{tot}}} \prod_i \left(1 + \frac{s_i}{b_i}\right)
\]

in this case. Here \(s_{\text{tot}}\) represents the total expected
signal over all signal sensitive data taking periods, i.e. a total
number of expected axion-induced X-rays recorded by our detector.

Further, systematics are taken into account by adding four 'nuisance
parameters', two for signal and background and two for the
position. We multiply the likelihood function with one normal
distribution for each nuisance parameter, which is normalized such
that \(\mathcal{N}(\theta = 0,~\sigma) = 1\). The signal and
background parameters are scaled by \(\theta\), such that a positive
\(\theta\) increases the parameter and a negative decreases it. At the
same time the normal distribution acts as a penalty term. To compute
the limit the explicit \(\theta\) dependencies must be removed, which
is done by marginalization, i.e. integrating them out

\[
\mathcal{L}_{M} = \iiiint_{-\infty}^{\infty} e^{-s'_{\text{tot}}} \cdot \prod_i \left(1 +
\frac{s_i''}{b_i'}\right) \cdot
\exp\left[-\frac{\theta_b^2}{2 \sigma_b^2} - \frac{\theta_s^2}{2 \sigma_s^2} -
\frac{\theta_x^2}{2 \sigma_{xy}^2} - \frac{\theta_y^2}{2 \sigma_{xy}^2} \right]\, \dd \theta_b \dd \theta_s \dd \theta_x \dd \theta_y
\]

with \(a' = a ( 1 + \theta_a )\) and \(a'' = a ( 1 + \theta_a ) ( 1 + \theta_x ) ( 1 +
\theta_y )\) where $a \in \{s, b\}$. This keeps the possibility of variance due to systematics
including the penalization embedded in the marginal likelihood, but
restores a single variable likelihood function with a well defined
single value for the \(95^{\text{th}}\) percentile.

The limit \(g'_{ae}\) is then defined by

\begin{equation}
  \label{eq:limit:obtain}
  0.95 = \frac{\int_{-\infty}^{g_{ae}'} \mathcal{L}(g_{ae}) \pi(g_{ae}) \, \dd g_{ae}}{\int_{-\infty}^{\infty} \mathcal{L}(g_{ae}) \pi(g_{ae}) \, \dd g_{ae}}
\end{equation}

where \(\pi(g_{ae})\) represents the prior probability distribution
for the axion-electron coupling, taken as uniform in $g_{ae} > 0$ and
zero elsewhere. It is computed from an empirical cumulative
distribution function.

However, the evaluation of a four-fold integral where the integrand is
expensive to evaluate numerically, is challenging due to the 'curse of
dimensionality'. As such the Metropolis-Hastings Markov Chain Monte
Carlo algorithm \cite{metropolis53_mcmc,hastings70_mcmc} is used to
evaluate the integrand efficiently only in those regions of the
parameter space where the integrand contributes to the integral.

\subsection{Signal and background in detail}
\label{sec:limit:sig_back}

With an unbinned likelihood approach we use the position on the center
GridPix and its energy as the parameters of interest. This means we
need to be able to evaluate the expected signal as well as the background
at the arbitrary positions and energies that each candidate may have.

The expected signal \(s_i\) can be expressed as
\begin{equation}
\label{eq:expected_signal}
s_i(g_{ae}) = f(g_{ae}, E_i) \cdot A \cdot t \cdot P_{a \rightarrow \gamma, \text{vacuum}} \cdot \epsilon(E_i) \cdot r(x_i, y_i)
\end{equation}
where \(f(g_{ae}, E_i)\) is the solar axion flux at energy \(E_i\) as
a function of \(g_{ae}\), the area of the magnet bore \(A\), the total
tracking time \(t\). \(\epsilon(E_i)\) describes the energy dependence
of the detection efficiency, which combines the telescope effective
area \(\epsilon_{\text{tel}}\), detector window transparency
\(\epsilon_{\text{window}}\), absorption probability of X-rays in the
gas \(\epsilon_{\text{gas}}\), veto efficiency
\(\epsilon_{\text{veto}}\) and software efficiency
\(\epsilon_{\text{S}}\),

\begin{equation}
  \label{eq:efficiencies}
\epsilon(E_i) = \epsilon_{\text{tel}}(E_i) \cdot \epsilon_{\text{window}}(E_i) \cdot \epsilon_{\text{gas}}(E_i) \cdot \epsilon_{\text{veto}} \cdot \epsilon_{\text{S}}.
\end{equation}

\(P_{a \rightarrow \gamma, \text{vacuum}}\) is the conversion
probability of the axion converting into a photon in the magnetic
field $B$ of length $L$, given by

\[
P_{a\rightarrow\gamma, \text{vacuum}} = \epsilon_0 \hbar c^3 \left( \frac{g_{a\gamma} B L}{2} \right)^2.
\]

Finally, \(r(x_i, y_i)\) is the distribution of the axion flux over
the detection area. This is computed using the raytracing simulation
mentioned in section~\ref{sec:detector} based on the expected
axion emission in the solar core and the X-ray telescope optics. This
raytracing result is shown as part of
figure~\ref{fig:data_unblinding:axion_candidates} as a colored, partially transparent
region.
\footnote{The exact numbers used are \(B = \SI{8.8}{T}\),
  \(L = \SI{9.26}{m}\), \(d_{\text{bore}} = \SI{43}{mm}\),
  \(t = \SI{160.38}{h}\).}

In order to achieve a smooth and continuous interpolation of the background
over the entire chip, we construct a background interpolation based on all
clusters found in the background dataset. This is done by defining \(b_i\)
as a function of candidate position and energy using

\[
b_i(x_i, y_i, E_i) = \frac{I(x_i, y_i, E_i)}{W(x_i, y_i, E_i)}
\]

where \(I\) is an intensity defined over clusters within a range \(R\) and
a normalization weight \(W\). The intensity is given by (for clarity without arguments)

\[
I(x, y, E) = \sum_{b \in \{ \mathcal{D}(\vec{x}_b, \vec{x}) \leq R \}}\mathcal{M}(\vec{x}_b, E_b)
  = \sum_{b \in \{ \mathcal{D}(\vec{x}_b, \vec{x}) \leq R \} } \exp \left[ -\frac{1}{2} \mathcal{D}^2 / \sigma^2 \right],
\]

where we introduce \(\mathcal{M}\) to refer to the measure we use and
\(\mathcal{D}\) to our metric:

\begin{equation*}
\mathcal{D}( (\vec{x}_1, E_1), (\vec{x}_2, E_2)) =
  \begin{cases}
    (\vec{x}_1 - \vec{x}_2)^2 \text{ if } |E_1 - E_2| \leq R \\
    \infty \text{ if } (\vec{x}_1 - \vec{x}_2)^2 > R^2 \\
    \infty \text{ if } |E_1 - E_2| > R
  \end{cases}
\end{equation*}

Finally, the normalization weight is the 'volume' of our measure
within the boundaries set by our metric \(\mathcal{D}\):

\[
W(x', y', E') = \int_{\mathcal{D}(\vec{x'}, \vec{x}) \leq R} \int_{E' - E_c}^{E' + E_c} \mathcal{M}(x', y')\, \dd x \dd y \dd E
\]

\begin{figure}[tp]
\begin{subfigure}{0.5\linewidth}
  \includegraphics[width=1.0\linewidth]{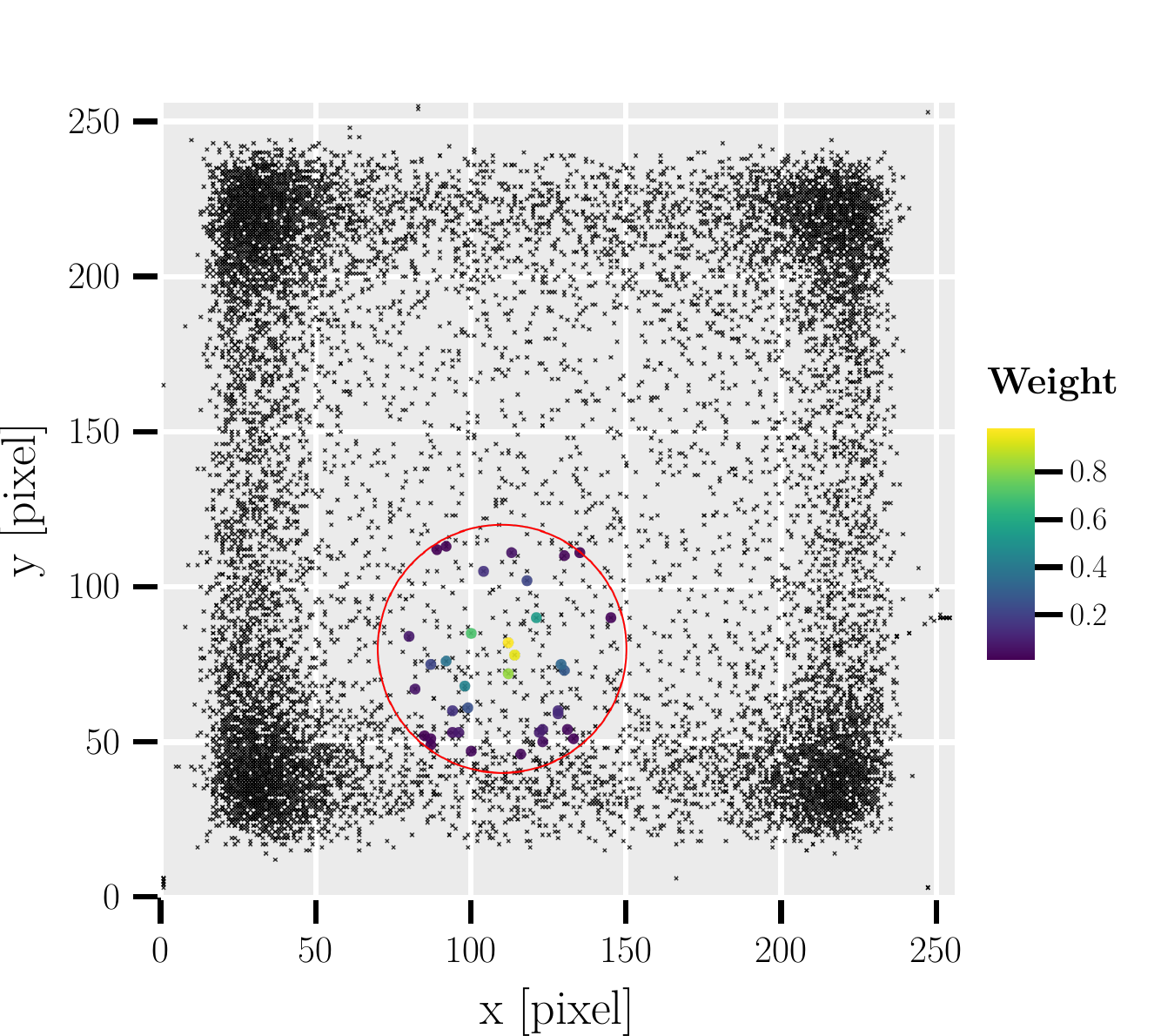}
  \caption{Intensity at a point}
  \label{fig:limit:interpolation_clusters}
\end{subfigure}%
\begin{subfigure}{0.5\linewidth}
  \includegraphics[width=1.0\linewidth]{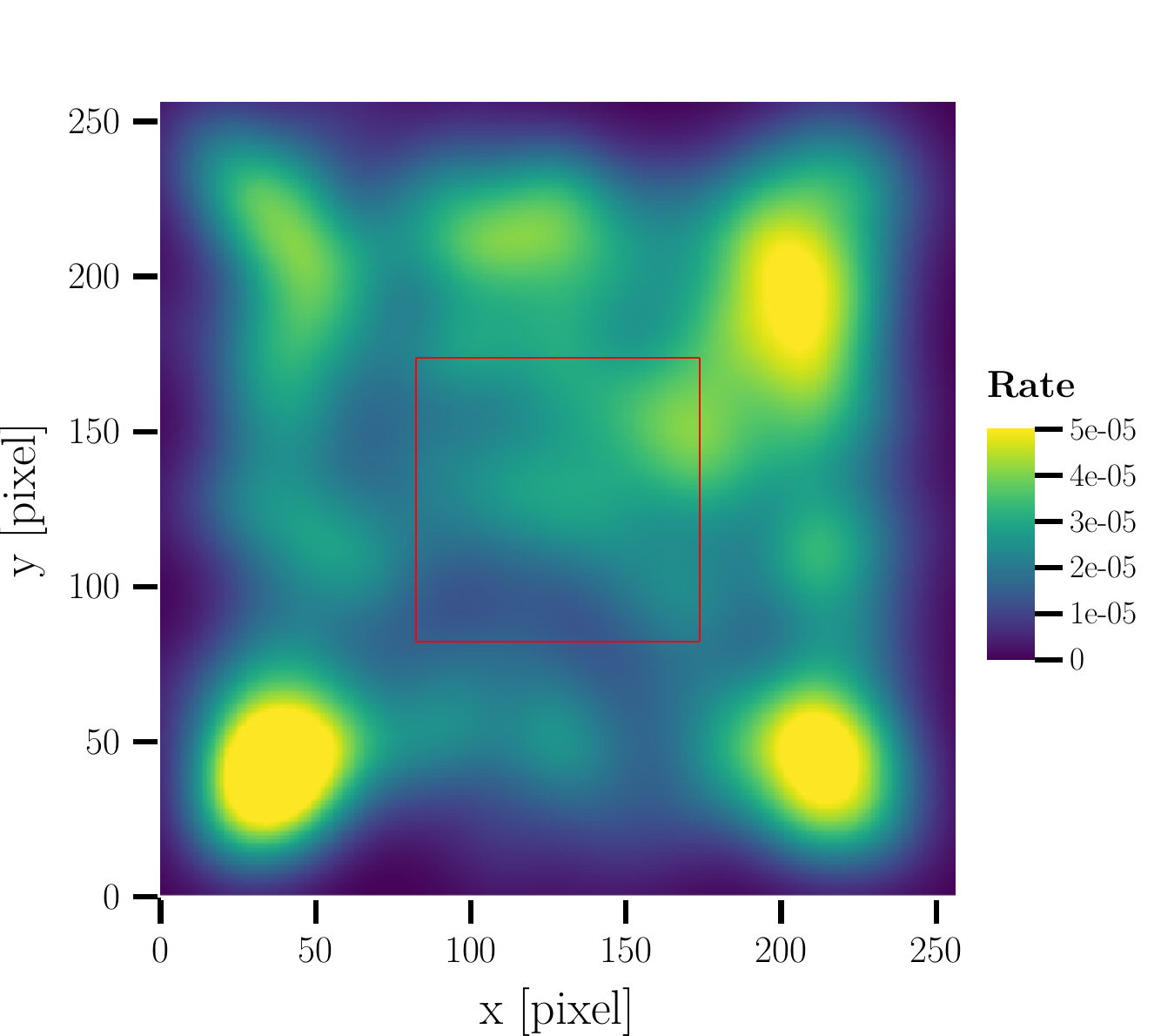}
  \caption{Interpolation}
  \label{fig:limit:background_interpolation_example}
\end{subfigure}%
\caption{\subref{fig:limit:interpolation_clusters}: Calculation of
  intensity $I$ at the center of the red circle. Black crosses
  indicate all background clusters. The red circle indicates cutoff
  $R$ in the x-y plane. Only clusters with colored dots inside the
  circle are within $\SIrange{2.70}{3.3}{keV}$. Their color is the
  weight based on the gaussian measure
  $\mathcal{M}$. \subref{fig:limit:background_interpolation_example}:
  Example of the resulting background interpolation at $\SI{3}{keV}$
  computed over the entire chip. A smooth, correctly normalized
  interpolation is obtained.}
\label{fig:limit:background_interp}
\end{figure}

The background clusters participating in the interpolation around a single
point in range $R$ and their associated weights for the interpolation
are shown in figure~\ref{fig:limit:interpolation_clusters}. This leads to the
anticipated interpolated background distribution, when applied to the whole
chip at a specific energy, as shown in
figure~\ref{fig:limit:background_interpolation_example}.
\footnote{Note, corrections need to be added towards the chip
edges where the radius of the metric is not fully contained on the
chip. This is done by upscaling the value within the chip by the
missing area.}

\subsection{Systematics}
\label{sec:limit:systmatics}
An overview on the systematics regarding the analysis can be found in ~\cite{schmidtPhD}. The combined values for the systematic uncertainty (including simulation) for the background ($\sigma_\text{b}$) the signal ($\sigma_\text{s}$) hypothesis, and the mechanical positioning of the detector ($\sigma_\text{xy}$) used for the limit calculation are:
\begin{align*}
\sigma_\text{s} &\leq \SI{3.38}{\percent} \text{ (assuming } \sigma_{\text{software}} = \SI{2}{\%} \text{)} \\
\sigma_\text{b} &= \SI{0.28}{\percent} \\
\sigma_\text{xy} &= \SI{5}{\percent} \text{ (fixed, uncertainty numbers are bounds)}.
\end{align*}

Note that the systematic of the signal rate $\sigma_\text{s}$ depends on the
exact choice of classifier and veto usage and hence the value there is
an upper bound assuming a worst-case software systematic contribution
of $\SI{2}{\%}$.

\subsection{Evaluation of parameters}
\label{sec:limit:eval_method}

As the limit calculation method is very sensitive to the combined
software efficiency, position and energy distribution of possible
candidates, the choice of parameters for the background suppression
methods are not fixed. Instead a variety of methods are evaluated and
expected limits are computed for each case. The set that yields the
best expected limit is the setup used to compute the real limit. Data
unblinding of the real tracking candidates was only performed after
the expected limits were computed.

An expected limit \(\langle L \rangle\) is defined by the median of sets
of toy candidate limits \(L_{t_i}\):

\[
\langle L \rangle = \mathrm{median}( \{ L_{t_i} \} ),
\]

where each limit is computed based on equation~\ref{eq:limit:obtain} using
a set of 'toy candidates' as inputs. Toy candidates are candidates
drawn from the background distribution. As our background model is an
interpolation in \((x, y, E)\), it is split into a grid of
\((10, 10, 20)\) grid cells. Each cell then contains an expected
(fractional) number of candidates if integrated over the tracking
time. This is used as the mean of a Poisson distribution to be sampled
from. The number of candidates sampled from said distribution is then
distributed over the entire grid cell volume. This produces one set of
toy candidates, for which a limit \(L_{t_i}\) is computed.

\begin{figure}[tbp]
\centering
\includegraphics[width=.9\linewidth]{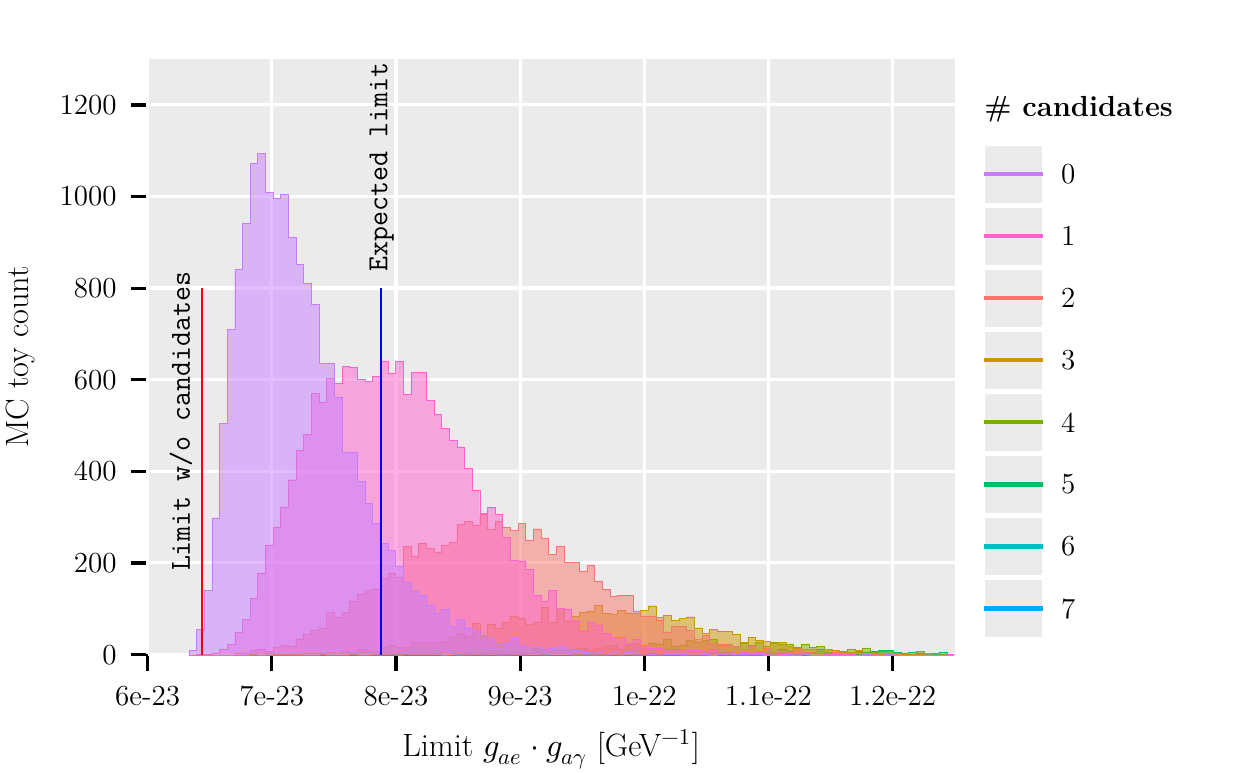}
\caption{\label{fig:mcmc_histo}Distribution of limits from
  \(\num{50000}\) toy candidate sets for the 'background only'
  hypothesis. The expected limit---the median---is shown as the blue
  line. The red line shows the limit for the case without any
  candidates. The different colored histograms correspond to toy sets
  with a different number of toy candidates in the signal sensitive
  region (not the \emph{total} number of candidates), defined by
  \(\ln(1 + s_i/b_i) > 0.5\). The most likely number of candidates in
  the sensitive region seems to be 0, 1 or 2.}
\end{figure}

Figure~\ref{fig:mcmc_histo} shows an example
for the determination of an expected limit for one set of
parameters. The blue line corresponds to the median of all computed
toy limits. The color of each histogram indicates how many candidates
were in the signal sensitive region (determined by \(\ln(1 +
\frac{s_i}{b_i}) > 0.5\)).

Table~\ref{tab:limit:eval_method} shows the best four parameter
options, which only differ by their software efficiency targeted by
the MLP \(\epsilon_\text{S}\), with the exception of the third row, which
uses no vetoes. Other considerations (e.g. different sets of vetoes)
produced worse limits.

\begin{table}[tbp]
\centering
\begin{tabular}{lrllrrr}
\toprule
Method             & \(\epsilon_\text{S}\) & Total eff. & Limit no signal [\(\si{GeV^{-1}}\)] & Expected limit [\(\si{GeV^{-1}}\)]\\[0pt]
\midrule
MLP                & \num{0.952(4)} & \num{0.797(7)} & \num{6.387e-23} & \num{7.878(6)e-23}\\[0pt]
MLP                & \num{0.980(3)} & \num{0.820(7)} & \num{6.20e-23} & \num{7.88(3)e-23}\\[0pt]
$\text{MLP}^{\dag}$ & \num{0.859(6)} & \num{0.859(6)} & \num{6.11e-23} & \num{7.95(5)e-23}\\[0pt]
MLP                & \num{0.906(5)} & \num{0.758(8)} & \num{6.47e-23} & \num{7.98(2)e-23}\\[0pt]
\bottomrule
\end{tabular}
\caption{\label{tab:limit:eval_method}Overview of the efficiencies and
  expected limits for the best performing four setups. \(\epsilon_\text{S}\):
  software efficiency of the MLP, 'Total eff.': \(\epsilon_\text{S} \cdot
  \epsilon_{\text{veto}}\). The first two
  entries and the last use the vetoes mentioned in the text, whereas
  the third entry $\text{MLP}^{\dag}$ is for the MLP only. Difference
  in uncertainty for first entry is due to larger number of samples.}
\end{table}

This implies a best expected limit \(g_{ae} \cdot g_{a\gamma} =
\SI{7.878(6)e-23}{GeV^{-1}}\) for an MLP software efficiency of
\(\SI{95}{\%}\) at a total combined efficiency of \(\SI{79.7(7)}{\%}\).

\section{Data unblinding and
  \texorpdfstring{\(g_{ae}\)}{\textit{g}\_{ae}} limit}
\label{sec:data_unblinding}

  \begin{figure}[tbp]
\centering
\includegraphics[width=0.9\textwidth]{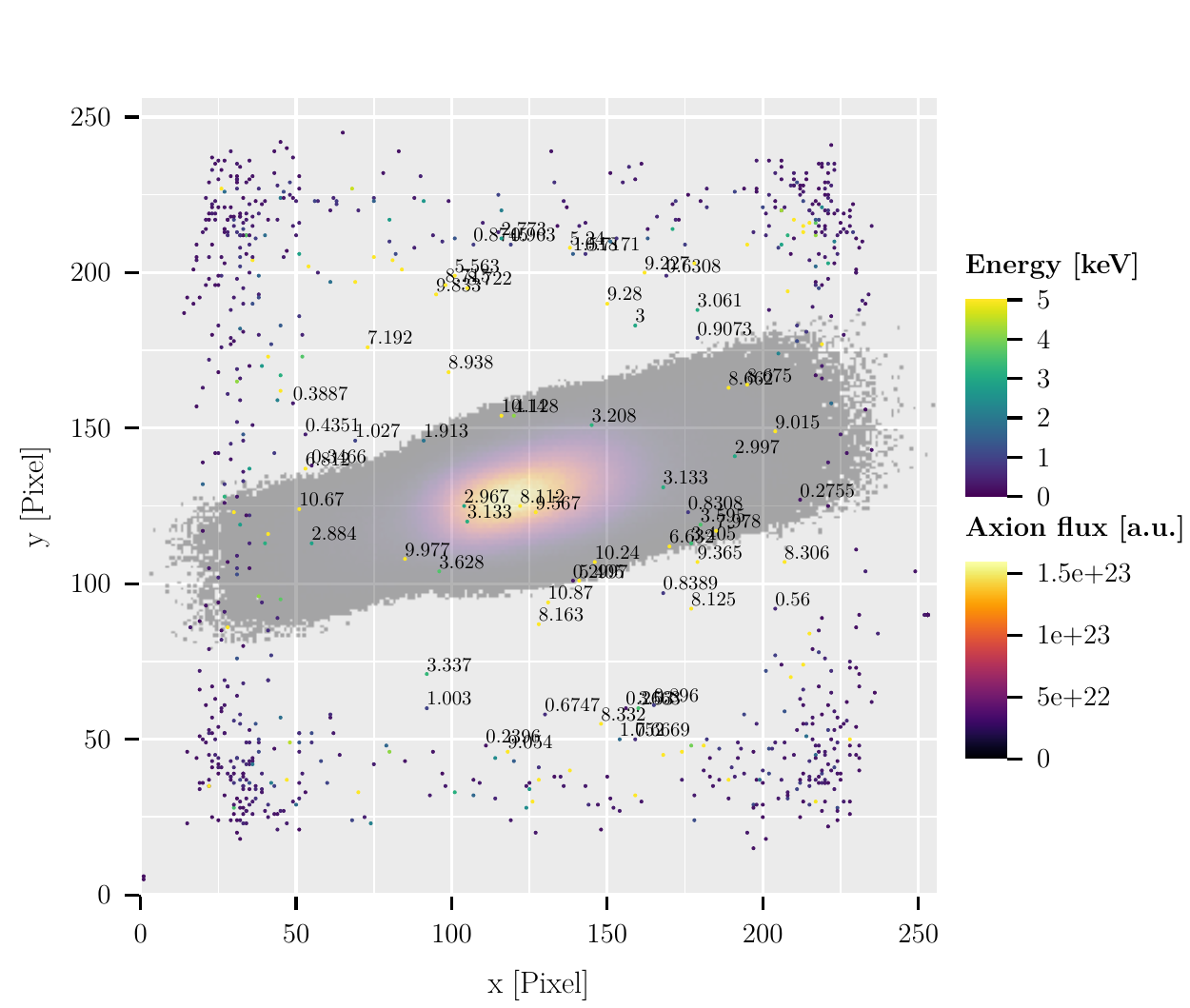}
\caption{\label{fig:data_unblinding:axion_candidates}Position and
  energy of all candidates obtained by application of setup with best
  expected limit. No candidates in signal and energy sensitive region
  are found.
  The axion image from a raytracing calculation is shown in the
  background as a colored, partially transparent region. It
  corresponds to the image produced assuming a Sun-Earth distance of
  $\SI{0.989}{AU}$ at \(\SI{0.3}{cm}\) behind the detector
  window. That means \(\SI{1.2}{cm}\) in front of focal point of the
  X-ray telescope at the median conversion point for expected axion
  induced X-ray flux. }
\end{figure}

\begin{figure}[tp]
\centering
\includegraphics[width=0.9\textwidth]{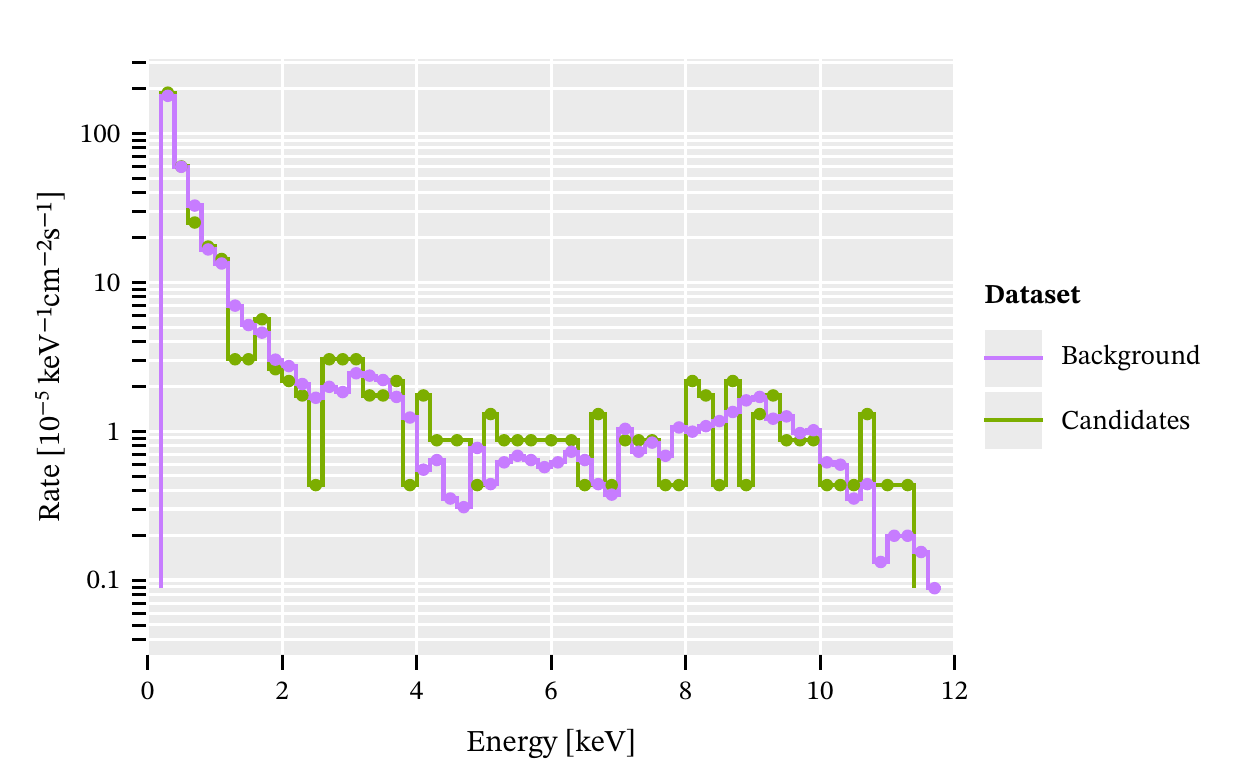}
\caption{\label{fig:data_unblinding:background_vs_signal} Comparison
  of the background and signal rate over the entire chip. No excess in
signal visible compared to background model.}
\end{figure}

With the expected limits present for different setups, data unblinding
was performed based on the setup with the best expected limit. The
candidates obtained by this are shown in figure~\ref{fig:data_unblinding:axion_candidates},
with the axion image underlaid and the energy of each candidate given
by the color as well as in text for those within a certain radius
around the center. As evident by the figure there are no relevant
candidates in terms of position (axion image location, hinted) and
energy (axion spectrum, see figure~\ref{fig:solar_axion_flux}) for an
axion signal (based on \(\ln(1 + s_i/b_i) < 0.5\) for each
candidate). This can also be seen in figure
\ref{fig:data_unblinding:background_vs_signal}, which shows the signal
rate against the background rate for the entire central chip. No
visible excess is seen.

The best published helioscope limit on the axion-electron coupling is
\cite{Barth_2013},

\[
  g_{ae} \cdot g_{a\gamma} < \SI{8.1e-23}{GeV^{-1}} \text{ at } \SI{95}{\%} \text{ CL}.
\]

In contrast the observed limit computed with the presented setup in
this work improves on this by $10\,\%$ to,

\[
\left(g_{ae} \cdot g_{a\gamma}\right)_{\text{observed}} < \SI{7.35e-23}{GeV^{-1}} \text{ at } \SI{95}{\%} \text{ CL}.
\]

\section{Summary and conclusion}
\label{sec:summary}
Based on the 7-GridPix detector operated at CAST in 2017 and 2018, the
limit on the axion-electron coupling could be improved to
\(\left(g_{ae} \cdot g_{a\gamma}\right)_{\text{observed}} <
\SI{7.35e-23}{GeV^{-1}}\) using approximately \(\SI{160}{h}\) of solar tracking
data. As the limit calculation method was designed to be generic under
the signal source, a limit on the axion-photon coupling constant
\(g_{a\gamma}\) was also computed to

\[
g_{a\gamma,~\text{observed}} < \SI{9.0e-11}{GeV^{-1}} \text{ at } \SI{95}{\%} \text{ CL}.
\]

While this is significantly higher than the previous best limit from
CAST \cite{cast_nature} limit at
\(g_{a\gamma,~\text{CAST}} < \SI{6.6e-11}{GeV^{-1}}\), this data has
already been used for a new combined limit with Micromegas data from
2019-2021 \cite{castcollaboration2024new} for the strongest helioscope
limit on the axion-photon coupling at \(g_{a\gamma,~\text{CAST,~2024}} < \SI{5.8e-11}{GeV^{-1}}\).
% \printbibliography[heading=bibintoc]

\bibliographystyle{JHEP}
\bibliography{references}

\end{document}